\documentclass[12pt]{article}
\usepackage[margin=2.5cm]{geometry}

\usepackage{amsfonts,amssymb,amsbsy,amsmath,amsthm,latexsym,natbib,graphicx}
\usepackage{subfigure,color,comment,multirow}

\theoremstyle{definition}

\newtheorem{remark}{Remark}

\newcommand{\calX}{\mathcal{X}}

\newcommand{\pihat}{\hat{\pi}}

\newcommand{\Zhat}{\hat{Z}}

\newcommand{\Lhat}{\hat{L}}

\newcommand{\Btil}{\tilde{B}}

\newcommand{\pitil}{\tilde{\pi}}

\newcommand{\Prob}[1]{\mathbb{P} \left({#1}\right)}

\newcommand{\Expect}[1]{\mathbb{E} \left[{#1}\right]}

\newcommand{\rtil}{\tilde{r}}

\newcommand{\Coffin}{\mathsf{C}}
\newcommand{\coffin}{\mathsf{c}}
\newcommand{\Region}{\mathcal{R}}

\newtheorem{example}{Examples}

\title{Exact Bayesian inference for discretely observed Markov Jump Processes using finite rate matrices}
\author{Chris Sherlock$^1$\footnote{c.sherlock@lancaster.ac.uk} and Andrew Golightly$^2$}
\date{{\small $^1$Department of Mathematics and Statistics, Lancaster University, UK\\
$^{2}$Department of Mathematical Sciences, Durham University, UK \\}}
\graphicspath{{./Graphics/}}

\begin{document}
\maketitle

\begin{abstract}
  We present new methodologies for Bayesian inference on the rate parameters of a discretely observed continuous-time Markov jump process with a countably infinite statespace. The usual method of choice for inference, particle Markov chain Monte Carlo (particle MCMC), struggles when the observation noise is small. We consider the most challenging regime of exact observations and provide two new methodologies for inference in this case: the minimal extended statespace algorithm (MESA) and the nearly minimal extended statespace algorithm (nMESA). By extending the Markov chain Monte Carlo statespace, both MESA and nMESA use the exponentiation of finite rate matrices to perform exact Bayesian inference on the Markov jump process even though its statespace is countably infinite. Numerical experiments show improvements over particle MCMC of between a factor of three and several orders of magnitude.
\end{abstract}

\noindent%
\vfill

\section{Introduction}
\label{sec.intro}

We consider exact inference for Markov jump processes (MJPs), continuous-time Markov chains which arise from reaction networks: stochastic models for the joint evolution of one or more populations of \emph{species}. These species may be biological species \cite[e.g.][]{Wilkinson2018stochastic}, animal species \cite[e.g.][]{drovandi16}, interacting groups of individuals at various stages of a disease \cite[e.g.][]{AnderssonBritton:2000}, or counts of sub-populations of alleles \cite[][]{Moran1958}, for example. The state of the system is encapsulated by the number of each species that is present, and in many cases this number is unbounded and, consequently, the statespace is countably infinite. The system evolves via a set of \emph{reactions} whose rates depend on the current state. Section \ref{sec.examples} describes three examples of reaction networks. The number of possible `next' states given the current state is bounded by the number of reactions, which is typically small; thus the infinitesimal generator of the process, which can be viewed as a countably infinite `matrix', is sparse. The methods which we develop in this article can be applied to any MJP, but they are particularly effective when the generator of the MJP is sparse.

The usual method of choice for inference on discretely observed MJPs with a countably infinite statespace is particle Markov chain Monte Carlo \cite[particle MCMC,][]{AndrieuDoucetHolenstein:2010} using a bootstrap particle filter \cite[e.g.][]{andrieu09,GolightlyWilkinson:2011,mckinley2014,Owen:2015,koblents2015,Wilkinson2018stochastic}. 
Paths from the prior distribution of the process are simulated and then resampled according to weights that depend on the likelihood of the next observation given the simulated path. Typically, as the precision of an observation increases, its compatibility with most of the paths plummets, leading to low weights; consequently, the efficiency of bootstrap particle MCMC decreases substantially. We consider the situation that is most challenging of all for a particle filter: when the observations are exact. Recently, paths proposed from alternative stochastic processes which take the next observation into account have successfully mitigated against this issue within particle MCMC \cite[]{GoliWilk15,GolSher2019}, albeit at an increased computational cost. The first of these provides the primary particle-filter comparator for the very different inference methodology, inspired by \cite{GHS17}, that we will introduce.

The likelihood for a discretely observed continuous-time Markov chain
with a large but \emph{finite} statespace and a rate matrix (or infinitesimal 
generator) $Q$ is the product of a set of transition probabilities,
each of which requires the evaluation of $v^\top e^{Qt}$ for an
inter-observation time $t$ and a non-negative vector $v$ representing
the state at an observation time. Fast algorithms for exactly this
calculation, some specifically designed for sparse $Q$, are available
\cite[e.g.][]{Sidje1998,SidjeStewart1999,Dubious,AMH11} and some are applicable even when the number of possible states, $d$, is in the tens of thousands; however, many processes of interest have a countably infinite number of states, and an exact, matrix-exponential approach might appear impossible for such systems. Refuting this conjecture,  \cite{GHS17} describes an ingenious pseudo-marginal MCMC algorithm \cite[]{AndrieuRoberts:2009} that uses random truncations \cite[e.g.][]{McLeish2011,GlynnRhee2014} and coffin states to explore the parameter posteriors for MJPs with infinite statespaces using exponentials of finite rate matrices. Unlike other pseudo-marginal algorithms which use random truncation \cite[e.g.][]{Lyneetal2015}, the algorithm of \cite{GHS17} is guaranteed to produce unbiased estimates of the likelihood which are non-negative. The algorithm, however, suffers from several issues: the most important arises from the need to use a proposal distribution for the truncation level (see Section \ref{sec.GHS}). 
As a result, in some examples the algorithm is less efficient than the most appropriate particle MCMC algorithm (see Section \ref{sec.numcomp}).

We describe the minimal extended statespace algorithm (MESA) and the nearly minimal extended statespace algorithm (nMESA), inspired by the key novel idea in \cite{GHS17}. These are fast and efficient algorithms for exact inference on  Markov jump processes with infinite statespaces through exponentiation of finite-dimensional rate matrices. Essentially, a sequence of nested regions is defined, $\emptyset=\Region_0\subset \Region_1\subseteq \Region_2\subseteq\dots$, with $\lim_{r\rightarrow \infty}\Region_r=\mathcal{X}$, the statespace of the MJP. The statespace of the MCMC Markov chain is then extended to include $\rtil$, the index of the smallest region that contains the MJP, and the corresponding extended posterior can be calculated using only finite rate matrices.  In the examples we investigate we find that MESA and nMESA are anything from a factor of three to several orders of magnitude more efficient than the most efficient particle MCMC algorithm.

We conclude this section with three motivating examples of reaction networks; these three examples will be used to benchmark our algorithms, the algorithm of \cite{GHS17} and particle MCMC in Section \ref{sec.numcomp}. In Section \ref{sec.existing} we describe the algorithm of \cite{GHS17}, separating out the key idea of nested regions which is shared by our algorithms. Section \ref{sec.newalgos} describes MESA and nMESA themselves, and the article concludes in Section \ref{sec.discuss} with a discussion. 

\subsection{Reaction network examples}
\label{sec.examples}
In a reaction network, the state vector, $X$, consists of the (non-negative) counts of one or more physical, chemical or biological species.  The state vector is piecewise constant over time, and updates only when a reaction occurs. For a given system, let there be $R$ possible reactions. The state update is deterministic given the current state and the particular reaction. Reaction $r$ ($r=1,\dots,R$) occurs according to a Poisson process with a rate of $h_r(X; \theta_r)$ for some function $h_r$ and unknown parameter $\theta_r$; \emph{i.e.}, for some small time interval $\delta t$, $\mathbb{P}(\mbox{reaction}~r~\mbox{occurs in}~(t,t+\delta t]|X=x)=h_r(x;\theta_r)\delta t+o(\delta t)$. For example, the Lotka-Volterra model, below, has $R=3$ reactions, the first of which changes the state vector $x=(x_1,x_2)$ to $(x_1-1,x_2)$ and occurs with a rate of $h_1(x;\theta_1)=\theta_1 x_1$.
  
To motivate the importance of inference on reaction networks we now present: the Lotka-Volterra predator-prey model, the Schl\"ogel model, which is one of the simplest bistable networks, and a simple model of gene auto-regulation in prokaryotes. We will perform inference on these reaction networks in our simulation study in Section \ref{sec.numcomp}.

\begin{example}
  \label{example.LV} \textbf{The Lotka-Volterra predator-prey model} \cite[e.g.][]{Boys2008}. Two species, predators, $\mathsf{pred}$, and prey, $\mathsf{prey}$, with counts of $X_1$ and $X_2$ respectively evolve and interact through the following three reactions (with associated rates):
  \begin{align*}
    \mathsf{Pred}&\stackrel{\theta_1X_1}{\longrightarrow}\emptyset\\
    \mathsf{Prey}&\stackrel{\theta_2X_2}{\longrightarrow}2\mathsf{Prey}\\
    \mathsf{Pred}+\mathsf{Prey}&\stackrel{\theta_3X_1X_2}{\longrightarrow}2\mathsf{Pred}.
    \end{align*}
\end{example}

\begin{example}
  \label{example.Sch} \textbf{The Schl\"ogel model} \cite[e.g.][]{VellelaQian09}  has two stable meta states, and the frequency of transitions between the meta states is much lower than the frequency of transitions between states within a single meta state. The interactions between the single species, whose frequency is $X$, and two chemical `pools', $\mathsf{A}$ and $\mathsf{B}$ are:
  \[
  \begin{array}{rclcrcl}
    \mathsf{A}+\mathsf{2X}&\stackrel{\theta_1X(X-1)/2}{\longrightarrow}&3\mathsf{X}&,&
    \mathsf{B}&\stackrel{\theta_3}{\longrightarrow}&\mathsf{X}\\

    \mathsf{3X}&\stackrel{\theta_2X(X-1)(X-2)/6}{\longrightarrow}&2\mathsf{X}+\mathsf{A}&,&
    \mathsf{X}&\stackrel{\theta_4X}{\longrightarrow}&\mathsf{B}.
  \end{array}
  \]
\end{example}

\begin{example}
  \label{example.AR} The \textbf{autoregulatory gene network} of \cite{GolightlyWilkinson:2005} models the production of $\mathsf{RNA}$ from $\mathsf{DNA}$ and of a protein $\mathsf{P}$ from $\mathsf{RNA}$, as well as the extinction of both $\mathsf{RNA}$ and $\mathsf{P}$, the reversible dimerisation of $\mathsf{P}$ and the reversible binding of the dimer, $\mathsf{P}_2$ to $\mathsf{DNA}$, where the binding inhibits production of $\mathsf{RNA}$. The total number of copies of $\mathsf{DNA}$, $G$, is fixed, and the reactions are:
  \[
  \begin{array}{rclcrcl}
    \mathsf{DNA}+\mathsf{P}_2&\stackrel{\theta_1(G-X_4)X_3}{\longrightarrow}&
    \mathsf{DNA}\cdot \mathsf{P}_2&,&
     \mathsf{DNA}&\stackrel{\theta_3(G- X_3)}{\longrightarrow}& \mathsf{DNA}+\mathsf{RNA},\\
     \mathsf{DNA}\cdot \mathsf{P}_2&\stackrel{\theta_2X_4}{\longrightarrow}&\mathsf{DNA}+\mathsf{P}_2&,&
     \mathsf{RNA}&\stackrel{\theta_4X_1}{\longrightarrow}&\mathsf{RNA}+\mathsf{P},\\
     2\mathsf{P}&\stackrel{\theta_5X_2(X_2-1)/2}{\longrightarrow}&\mathsf{P}_2&,&\mathsf{RNA}&\stackrel{\theta_7X_1}{\longrightarrow}&\emptyset,\\
     \mathsf{P}_2&\stackrel{\theta_6X_3}{\longrightarrow}&2\mathsf{P}&,&
     \mathsf{P}&\stackrel{\theta_8X_2}{\longrightarrow}&\emptyset,
  \end{array}
  \]
  where $X_1,\dots,X_4$ denote the counts of $\mathsf{RNA}$, $\mathsf{P}$, $\mathsf{P_2}$, and $\mathsf{DNA\cdot P_2}$ respectively.
\end{example}

The potentially countably infinite set of possible states of a reaction network can be placed in one-to-one correspondence with the non-negative integers.
The $i,j$th entry of the corresponding rate `matrix' $Q$ ($i\ne j$) is the rate for moving from state $i$ to state $j$. 

\subsection{Notation}
Throughout this article, a scalar operation applied to a vector means that the operation is applied to each element of the vector in turn, leading to a new vector, \emph{e.g.}, for the $d$-vector $\theta$, $\log \theta\equiv (\log \theta_1,\dots, \log \theta_d)^\top$. We denote the vector of $1$s by $1$. The symbol $0$ denotes the scalar $0$ or the vector or matrix of $0$s as dictated by the context.

There is a one-to-one correspondence between any vector state $x$, such as numbers of predators and prey in Example \ref{example.LV}, and the associated non-negative integer state, which we denote by $k(x)$. Throughout this article, for simplicity of presentation, we abuse notation by abbreviating the $(k(x),k(x'))$ element of a matrix $M$, strictly $[M]_{k(x),k(x')}$, to $[M]_{x,x'}$.

\section{Inference for MJPs with infinite statespaces using the rate matrix}
\label{sec.existing}
For simplicity of presentation we assume a known initial condition, $x_0$, though the methodology is trivially generalisable to an initial distribution. As in \cite{GHS17}, we then  
 consider the observation regime where particle filters typically struggle most: exact counts of all species are observed at discrete points in time, $t_1,t_2,\dots,t_n$; for simplicity we present the case where $t_i=it$ for some inter-observation interval $t$.

 Throughout this article, $\theta$ denotes the vector of positive reaction-rate parameters, and Bayesian inference is performed on $\psi:=\log \theta$, to which a prior density $\pi_0(\psi)$ is assigned.

For a finite-statespace Markov chain, whilst the rate matrix, $Q$, is the natural descriptor of the process, the likelihood for typical observation regimes involves the transition matrix, $e^{Qt}$, the $(i,j)$th element of which is exactly
$\Prob{X_t=j|X_0=i}$. 
By the Markov property, the likelihood for the exact observations
$x_1,\dots,x_n$ is then
\begin{equation}
\label{eqn.finite.full.like}
  L(\psi;x_{1:n})
=
\prod_{i=1}^n
[e^{Q(\psi)t}]_{x_{i-1},x_i}.
\end{equation}

The above likelihood is used within the algorithm of \cite{GHS17}, and within MESA and nMESA. All three algorithms share the same construction of nested regions which enables the use of \eqref{eqn.finite.full.like} and which we now describe.

\subsection{Set up for countably infinite statespaces}
\label{sec.setup.inf}
Let the MJP, $\{X_s\}_{s\ge 0}$, have a statespace of $\calX$, start from $X_0=x$ and be observed  precisely at time $t$: $X_t=x'$.
Consider an infinite sequence of regions, $\{\Region_r\}_{r=0}^\infty$ with
$\Region_0=\emptyset \subset \Region_1\subseteq \Region_2\subseteq \Region_3\dots$ and $\lim_{r\rightarrow \infty}\Region_r=\calX$; we permit equality so that the description also applies to MJPs with finite statespaces. Furthermore, $\Region_1$ should be chosen such that
$\Prob{\{X_{s}\}_{0\le s\le t}\in\Region_1\mid X_0=x,X_t=x'}>0$.
Let $d_r=|\Region_r|$ be the number of states in $\Region_r$.

Let $Q(\psi)$ be the infinitesimal generator for the MJP on $\calX$.
For finite $A,B\subseteq \calX$, let $Q(A,B)$ be
the submatrix of $Q$ that involves
transitions from $A$ to $B$, and  
let $Q_r$ be the
 be the rate matrix for transitions inside $\Region_r$ except that
it has an additional coffin state, $\Coffin$, which receives all transitions that, under $Q$, would exit $\Region_r$. Specifically
\[
Q_r:=
\left[
  \begin{array}{c|c}
      Q(\Region_r,\Region_r)&\coffin\\
      0&0\\
      \end{array}
  \right],
\]
where, here and henceforth, $\coffin$ denotes the scalar or column vector (as
appropriate) such that $\sum_{j=1}^{d_r+1}Q_{i,j}=0$ for each $i$. We will, in fact, have a different sequence of regions defined for each inter-observation interval. We will denote the $r$th region for the $i$th inter-observation interval by
$\Region_r^{(i)}$ and the associated transition matrix by $Q_r^{(i)}$. For clarity of exposition, we will often suppress the superscript $(i)$.

For each region $\Region_r\subseteq \calX$, there is a one-to-one map $k_r:\Region_r\rightarrow \{1,\dots,d_r\}$ and we add that $k_r:\Coffin\rightarrow d_r+1$.
Using the shorthand of $X$ for $\{X_s\}_{s=0}^t$, for $0\le t_1<t_2\le t$ we define:
\begin{eqnarray*}
  B(X;t_1,t_2,\Region)&:=&1\{X_s\in \Region~\forall s\in [t_1,t_2]\}~(r\ge 0).
\end{eqnarray*}
We set $\rtil(X;t_1,t_2)$ to be the unique index where $X_s\in \Region_{\rtil(X;t_1,t_2)}~\forall s\in
 [t_1,t_2]$ and $\exists~s\in[t_1,t_2]~\mbox{such that}~X_s\notin
 \Region_{\rtil(X;t_1,t_2)-1}$; \emph{i.e.}, the smallest index $r$ such that $B(x;t_1,t_2,\Region_r)=1$. 

\subsection{The method of \cite{GHS17}}
\label{sec.GHS}
In \cite{GHS17}, henceforth abbreviated to GHS17, the random-truncation method of \cite{McLeish2011} and \cite{GlynnRhee2014} leads to an unbiased estimator of the likelihood of a set of observations, which feeds into a pseudo-marginal MCMC algorithm \cite[]{AndrieuRoberts:2009} targetting the posterior
 $\pi(\psi)\propto \pi_0(\psi)L(\psi)$. Unlike other uses of random truncation within MCMC (e.g. \cite{Lyneetal2015}; see also \cite{JacobThiery2015}), however, the unbiased estimator of GHS17 can never be negative. We first briefly describe the random truncation method, before detailing the algorithm of GHS17.

Let $z_{0},z_1,\dots$ be a sequence, with $z:=\lim_{i\rightarrow \infty}z_i<\infty$. Let $R\in \{1,2,\dots,\}$ be sampled from some mass function. Then 
\[
\Zhat:=z_0+\sum_{j=1}^R \frac{z_j-z_{j-1}}{\Prob{R\ge j}}
\]
is an unbiased estimator of $z$. This is because, subject to the condition of Fubini's Theorem, the order of sum and expectation can be exchanged, so
\[
\Expect{\Zhat}
=
z_0+\sum_{j=1}^\infty \frac{(z_j-z_{j-1})}{\Prob{R\ge j}}\Expect{\mathsf{I}(R\ge j)},
\]
and the result follows from the telescoping sum.

At each iteration of the algorithm of GHS17, a value for $r$ is sampled at random from some discrete probability mass function $\{p(r)\}_{r=1}^\infty$.
In GHS17, $\Prob{R> r\mid R\ge r}:=q_r=aq_{r-1}$ for some $a<1$ and with $q_0=1$. Consequently,
\begin{equation}\label{eqn.GHS.lighttail}
  \Prob{R>r}=a^{r(r+1)/2}.
  \end{equation}

Since $[e^{Q_r(\psi)t}]_{x,x'}=\Prob{X_t=x',B(X;0,t,\Region_r)=1|X_0=x}$, 
$\lim_{r\rightarrow \infty}[e^{Q_r(\psi)t}]_{x_,x'}=L(\psi;x,x')$.
Thus, if $X_0=x$ and $X_t=x'$ are consecutive observations,
\begin{equation}
  \label{eqn.rand.trunc}
\Lhat(\psi;x,x',r)=\sum_{j=1}^{r} \frac{1}{\Prob{R\ge j}}
\left\{[e^{Q_j(\psi)t}]_{x_,x'}-[e^{Q_{j-1}(\psi)t}]_{x,x'}\right\},
\end{equation}
is a realisation from an unbiased estimator for the likelihood contribution $L(\psi;x,x')=[e^{Qt}]_{x_,x'}$ (here $[e^{Q_{0}t}]_{x,x'}=0$). 

One estimator of the form \eqref{eqn.rand.trunc}, with its own independently sampled $r$, is created for each inter-observation interval, and  
$\Lhat(\psi;r_{1:n}):=\prod_{i=1}^n\Lhat(\psi;x_{i-1},x_i,r_i)$
  then provides a realisation from an unbiased estimator for the full likelihood.

Given a current position $\psi=\log \theta$  and a set of region indices $r_{1:n}$ we have a realisation of an unbiased likelihood estimator, $\Lhat(\psi;r_{1:n})$. One iteration of the pseudo-marginal algorithm of GHS17 proceeds as follows: first, propose a new position from some density $q(\psi'|\psi)$ then sample $r_{1}',\dots,r_n'$ independently from the mass function $p(r)$ to obtain a realisation, $\Lhat(\psi';r'_{1:n})$ of an unbiased estimator for $L({\psi'})$. The pseudo-marginal Metropolis-Hastings acceptance probability for the proposal $(\psi',r'_{1:n})$ is then 
$\alpha(\psi,\psi')=1 \wedge  [\pi_0(\psi')\Lhat({\psi'};r_{1:n}')q(\psi|\psi')]/
     [\pi_0(\psi)\Lhat(\psi;r_{1:n})q(\psi'|\psi)]$.
The pseudo-marginal algorithm can be viewed as a Metropolis-Hastings Markov chain on $\psi$ and $r_{1:n}$ with a target distribution proportional to
\[
\pi_0(\psi)\Lhat(\psi;r_{1:n})\prod_{i=1}^np(r_i),
\]
and a proposal of $q(\psi'|\psi)\prod_{i=1}^np(r'_i)$.
Because $\Lhat(\psi;R_{1:n})$ is unbiased, integrating out all of the auxiliary variables from the target leaves $\pi(\psi)\propto \pi_0(\psi)L(\psi)$, as desired.

Typically, because they arise from a sequence of differences, likelihood estimates obtained via random-truncation might be negative and hence unusable \cite[\emph{e.g.}][]{Lyneetal2015,JacobThiery2015}. However, for observations $X_0=x$ and $X_t=x'$,
\begin{align}
\nonumber
  [e^{Q_r(\psi) t}]_{x,x'}-[e^{Q_{r-1}(\psi)t}]_{x,x'}
  &=
  \Prob{X_1=x',B(X;0,t,\Region_r)=1|X_0=x,\psi}\\
  &~~~-\Prob{X_1=x',B(X;0,t,\Region_{r-1})=1|X_0=x,\psi}\\
  &=  \Prob{X_1=x',\rtil(X;0,t)=r|X_0=x,\psi},
  \label{eqn.probdiff}
  \end{align}
  which is non-negative; so, by construction, a negative likelihood estimate is impossible.

  Although it can never be negative, the random truncation algorithm in \eqref{eqn.rand.trunc} suffers from several related problems. The proposal $p(r_i)$ should reflect the patterns in the terms in the sum in \eqref{eqn.rand.trunc}: if $\left\{[e^{Q_j(\psi)t}]_{x_,x'}-[e^{Q_{j-1}(\psi)t}]_{x,x'}\right\}/\Prob{R\ge j} \rightarrow \infty$ as $j\rightarrow \infty$
    then the unbiased estimate will be unstable and have a high, or even infinite, variance; if, on the other hand the ratio goes to zero then unnecessarily large regions will frequently be used, resulting in the exponentiation of unnecessarily large rate matrices with unnecessarily high rates, increasing the computational expense.

  GHS17 states that the distributional form of $p(r)$, was chosen partly since it describes the steady state of many simple queuing systems. However, the M/M/1 queue, for example, has a geometric stationary distribution \cite[e.g.][Ch.11]{grimmett2001probability}. From \eqref{eqn.GHS.lighttail} the tails of $p(r)$ are very light compared to geometric tails.
Even if a heavier-tailed proposal were used, however, there is no obvious choice for its shape, or reason to believe the shape would be consistent across inter-observation intervals. Further, some species might have a different spread than others, requiring differently shaped regions. Finally, this shape would typically depend on $\theta$, as exemplified in the following remark.

\begin{remark}
  \label{remark.thoughtexpt}
  Consider a Lotka-Volterra system with moderate variability between $X_0=x$ and $X_1=x'$, but now divide the true reaction rates by $1000$, which is equivalent to slowing down time by a factor of $1000$. The most likely paths would be those with close to a minimal number of events to get from $x$ to $x'$, so $\Prob{\rtil(X;0,1)=1}\approx 1$; on the other hand, a large increase in all of the rates would see an approximately quasi-stationary distribution for most of the time interval so larger regions would be more likely.
  \end{remark}

  We will reformulate the likelihood, creating an explicit extended statespace, and giving a different distribution for $r$ and $r'$; as a result, there is no division by $\Prob{R\ge r}$ and, indeed, no requirement for a generic proposal $p(r)$. The potential for different amounts of variation between species and across intervals is allowed for by letting the shape of the cuboidal regions  vary and for the nature of the cuboids themselves to vary between observations, all governed by two tuning parameters. Further, instead of requiring a random number, $R$ of matrix exponentials for each inter-observation interval, our algorithm requires just one.
  
\section{New Algorithms}
\label{sec.newalgos}
\begin{figure}
\begin{center}
  \includegraphics[scale=0.8,angle=0]{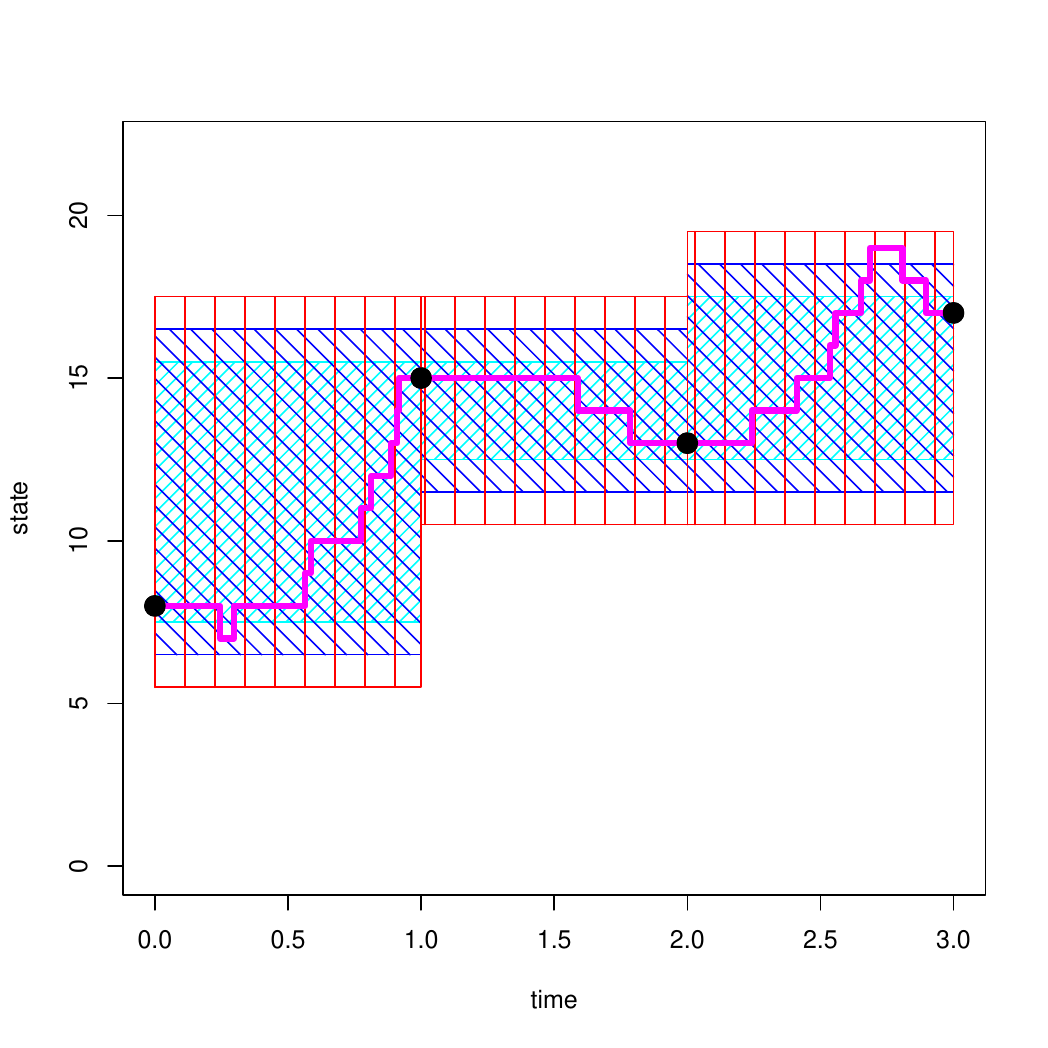}
\caption{Realisation (thick line) of $X$, a one-dimensional MJP, over the time interval $[0,3]$, with $x_0=8$, $x_1=15$, $x_2=12$ and $x_3=16$ (solid circles). For MESA, region 1 is the union of the three rectangles filled with diagonal increasing lines, and is contained within region $2$ which is the union of the three rectangles filled with diagonal decreasing lines, and is contained within and region three which is the union of the three rectangles filled with vertical lines. For nMESA, the same line types apply to the regions for each inter-observation interval. \label{fig.regions}
}
\end{center}
\end{figure}

We employ the same idea of a sequence of nested regions as in GHS17, with one sequence for each inter-observation interval. The \emph{nearly minimal extended statespace algorithm} (nMESA) has one auxiliary variable per interval as in GHS17, whereas the \emph{minimal extended statespace algorithm} (MESA) has a single auxiliary variable. In each case we explicitly extend the statespace from $\Psi$ to include the index
of the outermost region visited by $X$ over the observation window
(MESA) or between each pair of
observations (nMESA), and perform MCMC directly on this extended statespace. This partitioning of the space was defined in GHS17 but then unbiasedly integrated out via random truncation using auxiliary variables to set the truncation levels.
Figure \ref{fig.regions} depicts a realisation of an MJP together with the relevant regions for MESA and nMESA.
It provides a graphical representation of three regions (MESA, see Section \ref{sec.MESA}) or three regions per inter-observation interval (nMESA, see Section \ref{sec.nMESA}) together with a realisation of the MJP. 

\subsection{New regions}
\label{sec.new.regions}

For simplicity, each region, $\{\Region^{(i)}_r:i=1,\dots,n,r=1,\dots\}$,
is cuboid. Let the upper and lower bounds for species $s$ in region $r$ for inter-observation interval $i$ be $u_{r,s}^{(i)}$ and $l_{r,s}^{(i)}$; we refer $u_{r,s}^{(i)}-l_{r,s}^{(i)}+1$ as the \emph{width} of region $\Region^{(i)}_r$ for species $s$. In GHS17, for an interval between observations of $x$ and $x'$,  $\Region_1$ is the smallest cuboid that contains both $x$ and $x'$. However this cuboid does not necessarily allow a path between $x$ and $x'$. For example, since no reaction increases predator numbers by $1$, a Lotka-Volterra system with $x=(x_{1},x_{2})$ and $x'=(x_1+1,x_{2})$ must have left the rectangle with corners at $x$ and $x'$. We therefore define a scalar parameter $w_{\min}$, which specifies, for Region $\Region_1$, the smallest width for every species; if, for any species, the smallest cuboid that contains the two observations is narrower than $w_{\min}$ then the recursions below are performed for that species until this is no longer the case, leading to region $\Region_1$. Subsequent regions are obtained recursively from the previous region with the increase in region width for a given species proportional to the current width. For region $k+1$,
\begin{align*}
  u_{r+1,s}^{(i)}&= \mathsf{u}_s\wedge \left\{u_{r,s}^{(i)}+[1\vee \gamma (u_{r,s}^{(i)}-l_{r,s}^{(i)}+1)]\right\},\\
  l_{r+1,s}^{(i)}&= \mathsf{l}_s\vee \left\{l_{r,s}^{(i)}-[1\vee \gamma (u_{r,s}^{(i)}-l_{r,s}^{(i)}+1)]\right\}
\end{align*}
where, $\gamma>0$ is a tuning parameter and $\mathsf{l}_s$ and $\mathsf{u}_s$ are hard lower and upper bounds. Typically $\mathsf{l}_s=0$, whilst $\mathsf{u}_s$ is only required when there is an upper bound on the numbers for species $s$.
The regions in GHS17 use the above formulation, with $\gamma=w_{\min}=0$.

\subsection{The minimal extended statespace and target}
\label{sec.MESA}
For the observation regime given at the start of Section \ref{sec.existing}, define
\[
\Btil_r(X):=\prod_{i=1}^nB(X;t_{i-1},t_i,\Region_r^{(i)})~(r\ge 0).
\]
Thus $\Btil_r(X)=1$ if for every inter-observation interval, $X$ is entirely contained within that interval's region $r$. We denote the smallest index $r$ for which $\Btil_r(X)=1$ by $\rtil(X)$.
In every inter-observation interval the process is confined to that interval's region $\rtil(X)$ but in at least one inter-observation interval it is not confined to that interval's region $\rtil(X)-1$  ($\rtil=3$ in Figure \ref{fig.regions}). We target the extended posterior
\[
\pitil(\psi,r)\propto \pi_0(\psi) \Prob{x_{1:n},\rtil(X)=r\mid \psi,x_0},
\]
where
\[
\Prob{x_{1:n},\rtil(X)=r\mid \psi,x_0}
=
\prod_{i=1}^n \left[e^{Q_r(\psi)t}\right]_{x_{i-1},x_i}
-
\prod_{i=1}^n \left[e^{Q_{r-1}(\psi)t}\right]_{x_{i-1},x_i}.
\]
The marginal for $\psi$ is $\pi_0(\psi)\sum_{r=1}^\infty \Prob{x_{1:n},\rtil(X)=r\mid \psi,x_0}=
\pi_0(\psi)\Prob{x_{1:n}\mid \psi,x_0}$, as required.

\subsection{Nearly minimal extended statespace and target}
\label{sec.nMESA}
Let $\rtil_i(X)\equiv \rtil(X;t_{i-1},t_i)$ be, for the $i$th inter-observation interval, the index of the smallest region to
completely contain the MJP over that interval (in Figure \ref{fig.regions}, $(\rtil_1(X),\rtil_2(X),\rtil_3(X))=(2,1,3)$). We target the extended posterior
\[
\pitil(\psi,r_{1:n})\propto
\pi_0(\psi)\prod_{i=1}^n\Prob{x_i,\rtil_i(X)=r_i\mid \psi, x_{i-1}}
\]
where
\[
\Prob{x_{i},\rtil_i(X)=r_i\mid \psi,x_{i-1}}
=
  [e^{Q_{r_i}(\psi) t}]_{x_{i-1},x_{i}}-[e^{Q_{r_i-1}(\psi)t}]_{x_{i-1},x_{i}}.
\]
As for the MESA, the marginal for $\psi$ is the desired posterior, in this case since
\[
\sum_{r_{1:n}}\prod_{i=1}^n\Prob{x_i,\rtil_i(X)=r_i\mid \psi, x_{i-1}}
=\prod_{i=1}^n\sum_{r_i=1}^\infty\Prob{x_i,\rtil_i(X)=r_i\mid \psi, x_{i-1}}
=
\prod_{i=1}^n\Prob{x_i\mid \psi, x_{i-1}}.
\]

\subsection{The MCMC algorithms}
\label{sec.mcmc.algs}
Both MESA and nMESA use Metropolis-within-Gibbs algorithms. First, either $r\mid \psi,x_{0:n}$ (MESA) or $r_{1:n}\mid \psi,x_{0:n}$ (nMESA) is updated, then, respectively,  
$\psi \mid r,x_{0:n}$ or $\psi \mid r_{1:n},x_{0:n}$. For each algorithm, the $\psi$ update can, in principle, use any MCMC move that works on a continuous statespace; for simplicity and robustness we employ the random walk Metropolis. We propose $\psi'$ from a multivariate normal distribution centred on  $\psi$ and with a variance matrix proportional to the variance of $\psi$ obtained from an initial tuning run.
For MESA we update $r$ using a discrete random walk, proposing $r'=r-1$ or $r'=r+1$ each with a probability of $0.5$ (when $r=1$, the downward proposal is immediately rejected). For nMESA, conditional on $\psi$, each component, $r_i$, is updated independently via this symmetric discrete random-walk move. The random walk moves by a single region so as to save on computational cost. The new likelihood for a region $r'$ involves quantities of the form $v^\top [e^{Q_{r'}}]_{x_s,x_e}$ and $v^{\top}[e^{Q_{r'-1}}]_{x_s,x_e}$; when $r'$ is either $r+1$ or $r-1$, one of these quantities is already available from the likelihood calculations for the current region.

Both stages of both algorithms require the computation of the exponential of at least $n$ rate matrices. As with the algorithm of GHS17, therefore, our algorithm is, well suited to parallelisation if the rate matrices and/or the largest required matrix power are large; we do not pursue this here.

\subsection{Algorithm tuning and relative efficiency}
\label{SecAlgTuneEff}
We now consider the behavioural differences between MESA and nMESA and the effects of the tuning parameters.

MESA uses a single additional auxiliary variable, which specifies the region number for all inter-observation intervals, whereas nMESA has one variable per inter-observation interval. It is known \cite[e.g.][]{RobGelGilk1997} that many MCMC algorithms, including the random walk Metropolis mix more slowly on larger statespaces. This suggests that the MESA algorithm should be more efficient in terms of effective samples per iteration.

On the other hand, conditional on any given parameter vector $\psi$, consider the posterior distribution of the MJP $X$ over the set of of inter-observation intervals, and the varibility in the definition of the regions from one interval to the next.

Setting $w_{\min}$ to the minimum allowable for the system (e.g. $1$ for the Lotka-Volterra model and $0$ for the Schl\"ogel model) leads to the smallest $\Region_1$ sizes and the resulting matrix exponentials are very cheap to calculate. However, the larger the between-observation stochasticity the larger the region that is likely to be needed to contain the process between the observation times, yet some consecutive observations in the sequence will just happen to be similar, so the minimal $\Region_1$s will be too small. This is not an issue for nMESA which allows each inter-observation interval to find its own region level $\rtil_i$; but MESA forces all inter-observation intervals to use the \emph{same} region number, $\rtil$. The disparity between some of these $\Region_1$s very probably containing the process, and others very probably not containing it leads to poor mixing for MESA when $w_{\min}$ is at its minimum, and suggests increasing $w_{\min}$ to a sensible minimum value for any interobservation interval. Increasing $w_{\min}$ too far, for example beyond the range where the stochastic process is likely to lie, would lead to unnecessary computational expense, suggesting there may be an optimal $w_{\min}$ value.

Even with this larger $w_{\min}$, for many inter-observation intervals, the less flexible MESA will typically exponentiate larger matrices than nMESA. The numerical experiments in Section \ref{sec.numcomp} show that in the scenarios considered, nMESA's flexibility is more advantageous than MESA's smaller statespace, but typically by a factor of less than $2$. 

When the tuning parameter $\gamma=0$, for each species $\Region_{r+1}$ is $2$ units wider than the $\Region_{r}$. However, for regions of size $>>10$, say, this is a very small relative increase in width, and as such, we might expect an associated very small probability of the process staying within $\Region_{r+1}\backslash \Region_{r}$. In other words, the range of region indices over which the process is likely to need to move is large. Since in MESA and nMESA the MCMC move to change region proposes either an increase or decrease of the region index by $1$ (see Section \ref{sec.mcmc.algs}), region number mixes slowly. Since larger region indices are associated with larger reaction rates, for example, this also affects the mixing of $\psi$, all of which suggests choosing $\gamma>0$. On the other hand, consider a $\gamma$ so large that the process is almost certain to be in $\Region_1$; the dimension of $\Region_2$ will be approximately $(1+2\gamma)^{n_s}$ times the size of that of $\Region_1$, and may contain unnecessarily large rates, making matrix exponentials expensive to calculate, yet a move to $\Region_2$ is proposed every other iteration. These heuristics suggest that there should be an optimal $\gamma\in (0,\infty)$,

\section{Numerical comparisons}
\label{sec.numcomp}
\begin{figure}
\begin{center}
  \includegraphics[scale=0.30,angle=0]{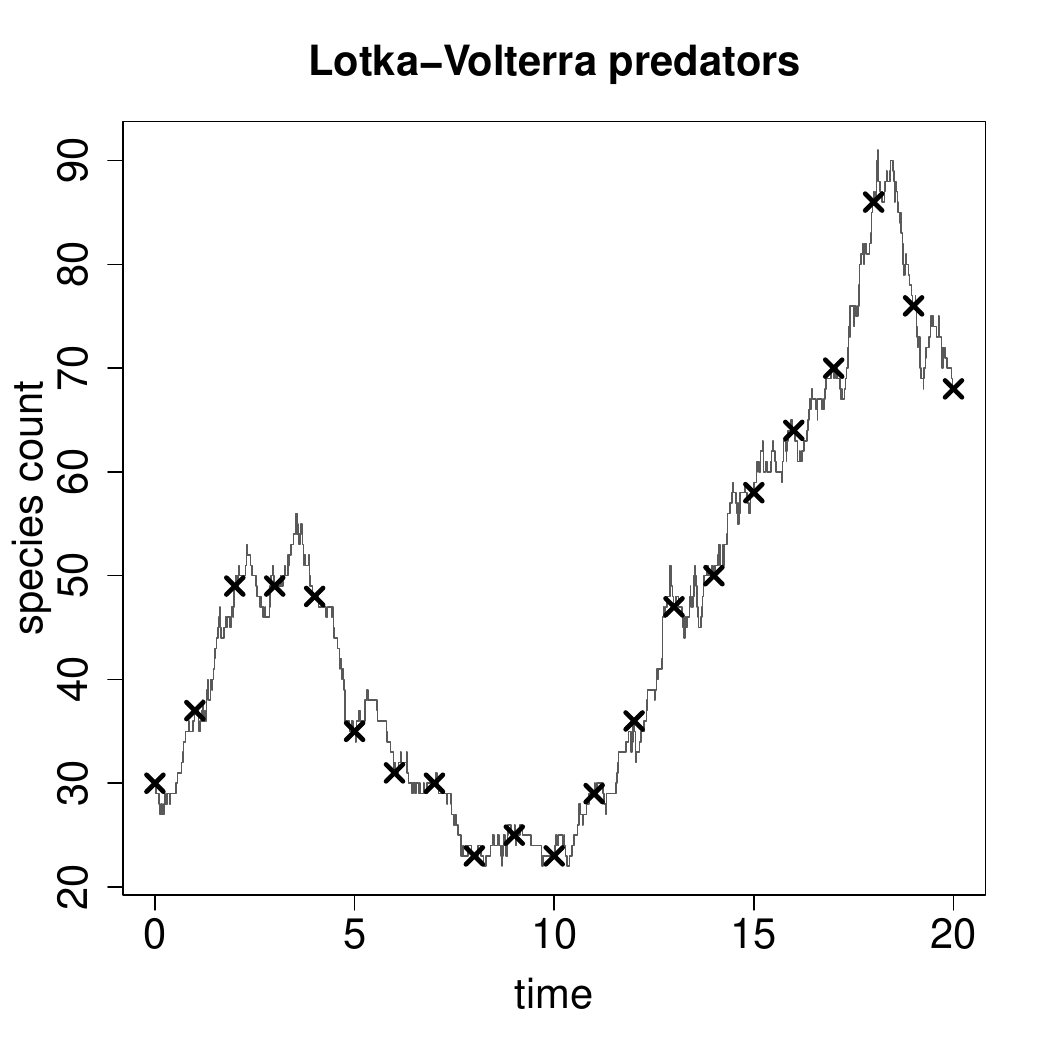}
  \includegraphics[scale=0.30,angle=0]{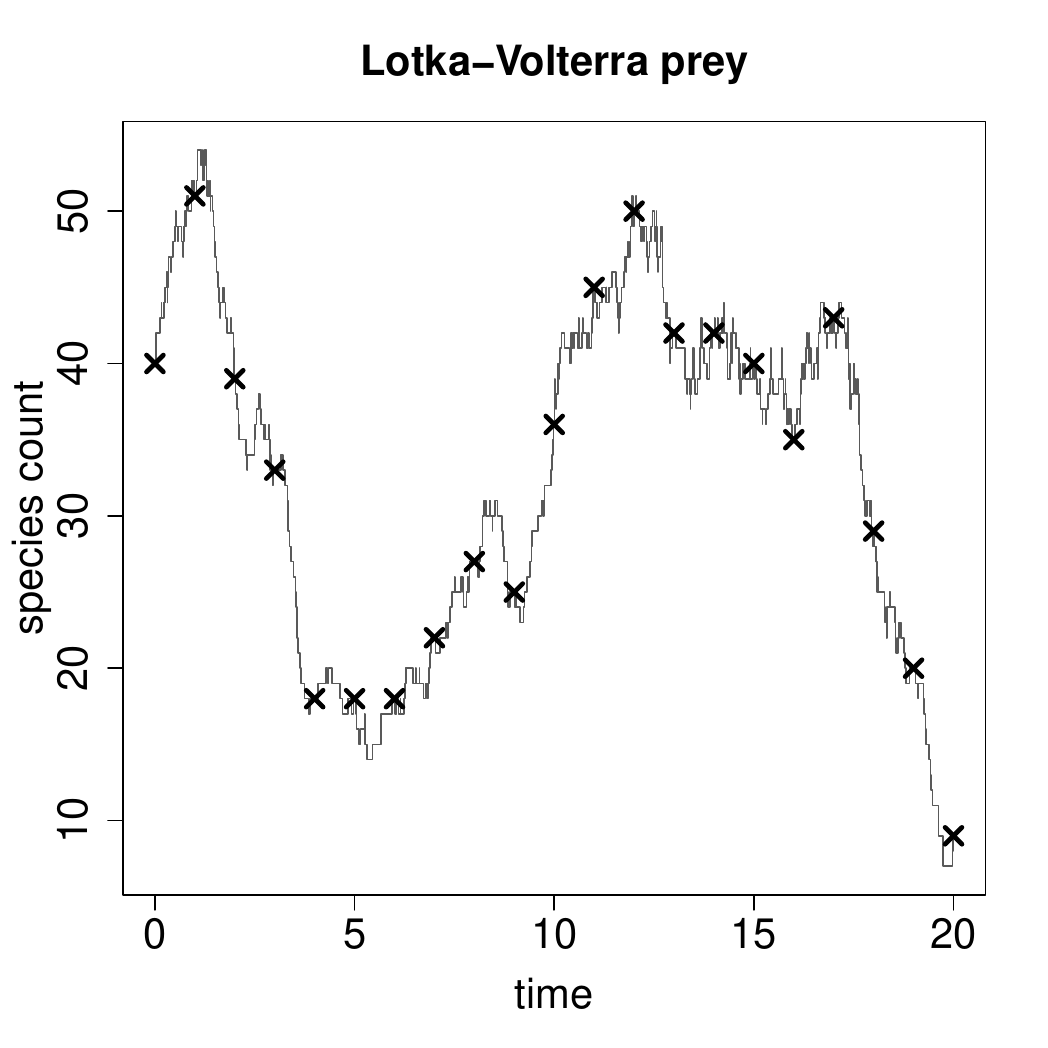}
  \includegraphics[scale=0.30,angle=0]{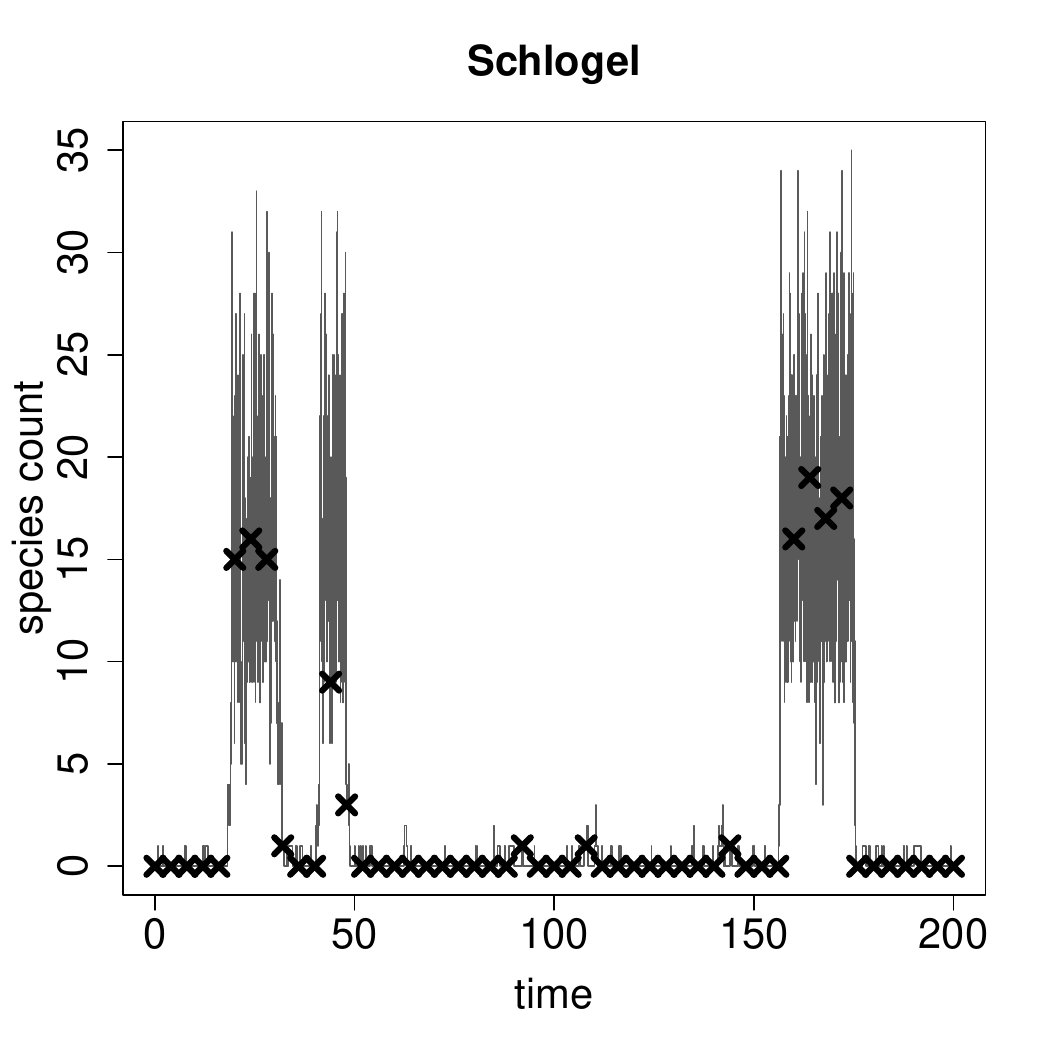}
\caption{Left (predators) and central (prey) panels: the realisation of the stochastic process from which datasets LV10, LV20 and LV40 arose, together with the LV20 data. Right panel: the Sch50 data with the Schl\"ogel process from which it arose. \label{fig.LVSchdata}
}
\end{center}
\end{figure}

For each of the reaction networks given in Section \ref{sec.examples} and a known initial condition, $x_0$, we simulated a realisation from the stochastic process from time $0$ to an appropriate $t_{end}$ and then recorded the states at regular intervals so that there were $50$ observations each from a realisation of the Schl\"ogel process and a realisation of the autoregulatory process, and $20$ observations of a Lotka-Volterra process; we name these data sets Sch50, AR50 and LV20, respectively. To investigate the effect of altering the inter-observation interval, for the Lotka-Volterra process, two further data sets, LV40 and LV10, were generated with $40$ and $10$ observations, respectively. To suppress the effect of inter-realisation variability, LV40 and LV10 were generated from the \emph{same} realisation as LV20 by, respectively, halving and doubling the inter-observation time interval; thus $LV10 \subset LV20 \subset LV40$. 
Figure \ref{fig.LVSchdata} shows the realisations of the stochastic processes from which LV20, LV40, LV10 and Sch50 arose, together with the observations in LV20 and Sch50. The realisation from the autoregulatory process and the observations AR50 are provided in Figure \ref{fig.ARdata} in Appendix \ref{sec.paramvals}, which also provides values for $x_0$, $t_{end}$ and the true parameters for all three processes (see Table \ref{table.exactinfo}) as well as the prior distributions assigned to the parameters.

For each data set the output from a tuning run of $10^4$ iterations of nMESA was used to create an estimate, $\widehat{\Sigma}$, of the variance matrix of the parameter vector, $\psi$. For comparability, for all algorithms, proposals for the random walk on $\psi$ were of the form: $\psi'=\psi +\lambda {\widehat{\Sigma}}^{1/2}z$, where $z$ is a realisation of a vector of standard Gaussians, $\widehat{\Sigma}^{1/2}$ is defined so that ${\widehat{\Sigma}^{1/2}}{\widehat{\Sigma}^{1/2\top}}=I$, and $\lambda$ is a tuning parameter.
The scaling, $\lambda$, of the random walk proposal was chosen using standard acceptance-rate heuristics \cite[e.g.][]{RobGelGilk1997,SFR2010,STRR2015}. The number of particles was chosen so that the variance of $\log \pihat$ at points in the main posterior mass was not much above $1$ \cite[][]{STRR2015,DDPK2015}.
No tuning advice is given in GHS17 so we proceeded by first tuning $\gamma$ and $a$ for a fixed, sensible $\psi$, and then tuning $\lambda$; see Appendix \ref{sec.tuningGHS} for further details.
Unless otherwise stated, each algorithm was run for $10^5$ iterations. In all cases, the first $100$ iterations were removed as burn in, as trace plots showed that this was all that was necessary. 

Results for MESA are presented in terms of the CPU time in seconds, $T$, the acceptance rate for the random walk update on the parameters, $\alpha_{\psi}$, the acceptance rate for the integer random walk update on the region, $\alpha_{r}$, and the number of effective samples per minute (rounded to the nearest integer) for the parameters, the region index and $\log \pi$. The number of effective samples was calculated using \texttt{effectiveSize} command in the \texttt{coda} package in \texttt{R} \cite[]{CODA}. Quantities are the same for nMESA except that $\alpha_{r}$ is the mean of the acceptance rates for the random walks on each of the region indices, and the effective samples per minute of the average region index is recorded. Neither particle MCMC nor GHS17 has a `region' auxiliary variable, so the two fields for this are left blank. GHS17 strictly should have $\gamma=0$, but we also report the results for larger $\gamma$ where this did not reduce the efficiency too much.

\subsection{Numerical and computational issues}
\label{sec.mxexp}
Calculations of the form $v^\top e^{Qt}$ were performed using the more efficient of two possible algorithms, chosen automatically at runtime on a case-by-case basis. The uniformisation method \cite[e.g.][]{SidjeStewart1999} and a variation on the scaling and squaring method \cite[e.g.][]{Dubious}. In our examples, $Q$ is a sparse $d\times d$ matrix with $\mathcal{O}(d)$ non-zero entries; define $\rho:=\max_{i=1,\dots d}|Q_{ii}|$.  With this set up, the uniformisation method takes $\mathcal{O}(\rho td)$ operations and has a memory footprint of $\mathcal{O}(d)$, whereas the scaling and squaring method has takes $\mathcal{O}(d^3 \log \rho)$ operations with a footprint of $\mathcal{O}(d^2)$. The latter was typically only used for some of the calculations for the Schlo\"gel model where for some of the observation intervals, with the MESA and GHS17 algorithms, typically, $\rho\gtrapprox 10^8$ but $d\lessapprox 10^3$. See Appendix \ref{sec.matexpdetail} for more details on the methods. 

The maximum size of an \texttt{unsigned integer} in \texttt{C++} is $\approx 4\times 10^9$. Both methods of exponentiation require the evaluation of $v^\top M^j$ for some matrix, $M$ with no negative entries, for some integer power $j\sim\rho+\mathcal{O}(\sqrt{\rho})$. Straightforward, exact (to a prescribed small tolerance) evaluation requires storing $j$ as an integer. For the Schl\"ogel system using GHS17 or using MESA with small $w_{\min}$, for some inter-observation intervals on some iterations $\rho>4\times 10^9$, sometimes considerably so. In such cases $\rho$ was truncated to $4\times 10^9$ so that inference was no longer exact, but could at least continue; in the remainder of this section we refer to this as the \emph{integer overflow} problem. With an increase in code and algorithm complexity this issue could be overcome, but on the occasions when it occurred the algorithms for which it occurred, even with the inexact inference, were much less efficient than MESA with a larger $w_{\min}$ or nMESA,  so we did not pursue this further.

Unless stated otherwise runs were performed on a desktop machine using a single thread of a single i7-3770 core. Code for MESA, nMESA and GHS17 is available from \url{https://chrisgsherlock.github.io/Research/publications.html} .

\subsection{Lotka-Volterra model}

Tables \ref{table.LV20}, \ref{table.LV40} and \ref{table.LV10} show the simulation results for a selection of the runs performed, respectively, for the LV20, LV40 and LV10 data sets. We focus on the LV20 data set, and point out any differences evident in the other two data sets.

Firstly, particle MCMC using the bridge of \cite{GoliWilk15}, henceforth referred to as GW15, was approximately a factor of $4$ times as efficient as the algorithm of GHS17 (factors of approximately $3$ and $7$ for LV40 and LV10 respectively). Further, runs of GHS17 with $\gamma\approx 0.1$ (best performance was for LV10, which is shown) were much less efficient than with $\gamma=0.0$. The most efficient MESA tuning was a factor of $3$ more efficient than particle MCMC (factors of approximately $7$ and $2.5$ for LV40 and LV10), whilst the most efficient nMESA was a factor of nearly $5$ times more efficient than particle MCMC (factors of approximately $12$ and $4.5$ for LV40 and LV10).

\begin{table}
\begin{center}
  \caption{Tuning parameter settings and results for GHS17, MESA and nMESA for the LV20 dataset. $^1$GW15 used 100 particles. $^2$GHS17 used $a=0.98$.\label{table.LV20}}
  
\begin{tabular}{l|rrr|rrr|rrrrr}
  \multicolumn{7}{c}{}&\multicolumn{5}{c}{ESS/minute}\\
  Algorithm&$\lambda$&$\gamma$&$w_{\min}$&$T$&$\alpha_{\psi}$&$\alpha_{r}$&
  $\psi_1$&$\psi_2$&$\psi_3$&$r$ or $\bar{r}$&$\log \pi$\\
  \hline
  GW15$^1$&$1.6$&&&$8157$&$15.5$&&$24$&$25$&$26$&&$10$\\
  GHS17$^2$&$1.6$&$0.0$&$1$&$17874$&$6.3$&&$6$&$6$&$6$&&$4$\\
  \hline
  MESA&$1.5$&$0.0$&$1$&$8270$&$28.2$&$73.5$&$64$&$65$&$62$&$42$&$58$\\
  MESA&$1.5$&$0.1$&$1$&$12634$&$28.4$&$72.2$&$42$&$42$&$42$&$36$&$42$\\
  MESA&$1.5$&$0.1$&$7$&$8023$&$28.2$&$68.9$&$65$&$65$&$64$&$85$&$66$\\
  MESA&$1.5$&$0.0$&$10$&$7579$&$28.2$&$70.8$&$68$&$69$&$68$&$54$&$66$\\
  MESA&$1.5$&$0.1$&$10$&$8035$&$28.2$&$67.1$&$67$&$68$&$69$&$95$&$72$\\
MESA&$1.5$&$0.1$&$20$&$7557$&$28.2$&$51.4$&$72$&$73$&$71$&$147$&$74$\\
MESA&$1.5$&$0.1$&$30$&$9396$&$28.9$&$1.3$&$62$&$64$&$62$&$194$&$61$\\
\hline
  nMESA  &$1.4$&$0.0$&$1$&$2887$&$27.8$ &$69.4$&$106$&$111$&$105$&$78$&$140$\\
  nMESA&$1.4$&$0.1$&$1$&$2991$&$27.7$ &$69.2$&$106$&$105$&$104$&$78$&$157$\\
  nMESA  &$1.4$&$0.2$&$1$&$4164$&$27.6$ &$59.5$&$81$&$85$&$84$&$90$&$264$\\
  nMESA  &$1.4$&$0.3$&$1$&$5553$&$27.9$ &$50.2$&$69$&$70$&$71$&$91$&$252$\\
  nMESA&$1.4$&$0.0$&$10$&$3054$&$28.8$&$53.6$&$115$&$117$&$115$&$75$&$93$\\
  nMESA&$1.4$&$0.1$&$10$&$3126$&$28.7$&$53.6$&$116$&$117$&$115$&$89$&$115$\\
  nMESA&$1.4$&$0.2$&$10$&$3745$&$28.9$&$40.9$&$114$&$114$&$108$&$276$&$180$\\
  nMESA&$1.4$&$0.1$&$20$&$4458$&$30.6$&$4.2$&$120$&$119$&$118$&$192$&$171$\\
  nMESA&$1.4$&$0.1$&$30$&$9480$&$31.6$&$0.07$&$63$&$62$&$60$&$177$&$65$\\
\end{tabular}
\end{center}
\end{table}

\begin{table}
\begin{center}
  \caption{Tuning parameter settings and results for GHS17, MESA and nMESA for the LV40 dataset. $^1$GW15 used 170 particles. $^2$GHS17 used $a=0.98$.\label{table.LV40}}
\begin{tabular}{l|rrr|rrr|rrrrr}
  %
  \multicolumn{7}{c}{}&\multicolumn{5}{c}{ESS/min}\\
  Algorithm&$\lambda$&$\gamma$&$w_{\min}$&$T$&$\alpha_{\psi}$&$\alpha_{r}$&
  $\psi_1$&$\psi_2$&$\psi_3$&$r$ or $\bar{r}$&$\log \pi$\\
  \hline
  GW15$^1$&$1.5$&&&$14039$&$16.9$&&$15$&$14$&$15$&&$6$\\
  GHS17$^2$&$1.4$&$0.0$&&$19320$&$6.7$&&$5$&$6$&$5$&&$5$\\
\hline
MESA&$1.5$&$0.0$&$1$&$5252$&$27.5$&$59.2$&$97$&$101$&$101$&$145$&$105$\\
MESA&$1.5$&$0.1$&$1$&$5372$&$27.7$&$59.3$&$103$&$102$&$99$&$131$&$103$\\
MESA&$1.5$&$0.0$&$7$&$4964$&$27.5$&$57.3$&$105$&$106$&$106$&$151$&$111$\\  
MESA&$1.5$&$0.1$&$7$&$5073$&$27.6$&$57.3$&$98$&$108$&$112$&$148$&$112$\\
  \hline
nMESA  &$1.4$&$0.0$&$1$&$1672$&$25.6$ &$59.1$&$164$&$162$&$161$&$155$&$312$\\
nMESA  &$1.4$&$0.1$&$1$&$1680$&$25.7$ &$59.2$&$169$&$170$&$169$&$162$&$317$\\
nMESA&$1.4$&$0.0$&$7$&$1805$&$27.3$&$37.8$&$186$&$196$&$202$&$196$&$220$\\
nMESA&$1.4$&$0.1$&$7$&$1807$&$27.3$&$37.9$&$180$&$185$&$186$&$179$&$203$
\end{tabular}
\end{center}
\end{table}

\begin{table}
\begin{center}
  \caption{Tuning parameter settings and results for GW15, GHS17, MESA and nMESA for the LV10 data set. $^1$GW15 used $140$ particles. $^2$GHS17 used $a=0.98$; with $\gamma=0.1$, GHS17 was run for $10^4$ iterations. \label{table.LV10}}
\begin{tabular}{l|rrr|rrr|rrrrr}
  \multicolumn{7}{c}{}&\multicolumn{5}{c}{ESS/min}\\
  Algorithm&$\lambda$&$\gamma$&$w_{\min}$&$T$&$\alpha_{\psi}$&$\alpha_{r}$&
  $\psi_1$&$\psi_2$&$\psi_3$&$r$ or $\bar{r}$&$\log \pi$\\
  \hline
  GW15$^1$&$1.6$&&&$10324$&$14.6$&&$14$&$16$&$17$&&$6$\\
  GHS17$^2$&$1.4$&$0.0$&&$19800$&$3.6$& &$2$&$3$&$2$&&$2$\\
  GHS17$^2$&$1.4$&$0.1$&&$10773$&$4.1$& &$0.6$&$0.5$&$0.7$&&$0.5$\\
  \hline
  MESA &$1.5$&$0.0$&$1$&$18615$&$27.9$&$85.9$&$24$&$25$&$29$&$7$&$20$\\
  MESA &$1.5$&$0.1$&$1$&$56925$&$27.9$&$72.4$&$9$&$9$&$9$&$7$&$9$\\
  MESA &$1.5$&$0.1$&$7$&$33932$&$27.9$&$72.3$&$16$&$15$&$16$&$14$&$16$\\
  MESA &$1.5$&$0.1$&$14$&$18172$&$27.7$&$68.1$&$27$&$28$&$27$&$34$&$31$\\
  MESA &$1.5$&$0.0$&$20$&$14274$&$28.1$&$82.6$&$32$&$34$&$34$&$12$&$25$\\
  MESA &$1.5$&$0.1$&$20$&$14911$&$27.9$&$65.2$&$36$&$36$&$35$&$53$&$37$\\
  MESA &$1.5$&$0.1$&$30$&$13144$&$27.8$&$51.7$&$41$&$40$&$41$&$82$&$44$\\
  MESA &$1.5$&$0.1$&$40$&$14659$&$28.1$&$6.4$&$36$&$37$&$37$&$90$&$38$\\
  \hline
    nMESA  &$1.4$&$0.0$&$1$&$6355$&$28.7$&$79.7$&$49$&$53$&$58$&$19$&$37$\\
    nMESA  &$1.4$&$0.1$&$1$&$12244$&$28.7$&$75.1$&$29$&$30$&$31$&$15$&$45$\\
    nMESA  &$1.4$&$0.2$&$1$&$21082$&$28.9$&$62.1$&$20$&$20$&$21$&$24$&$44$\\
  nMESA  &$1.4$&$0.1$&$14$&$7570$&$29.3$&$60.0$&$48$&$52$&$57$&$37$&$78$\\
  nMESA&$1.4$&$0.0$&$20$&$6132$&$29.4$&$43.2$&$54$&$56$&$58$&$23$&$36$\\
    nMESA&$1.4$&$0.1$&$20$&$6469$&$29.4$&$37.6$&$65$&$71$&$67$&$81$&$94$\\
  nMESA&$1.5$&$0.2$&$20$&$7857$&$29.7$&$29.4$&$62$&$63$&$66$&$148$&$126$\\
    nMESA&$1.5$&$0.1$&$30$&$8177$&$27.6$&$7.1$&$60$&$63$&$63$&$96$&$86$\\
    nMESA&$1.5$&$0.1$&$40$&$13652$&$28.1$&$0.7$&$40$&$40$&$40$&$83$&$40$\\
\end{tabular}
\end{center}
\end{table}

For MESA and nMESA, the variations in efficiency with $\gamma$ and many of the variations with $\gamma$ visible in Tables \ref{table.LV20}, \ref{table.LV40} and \ref{table.LV10} are explained by the heuristics in Section \ref{SecAlgTuneEff}. Further, any trend (or drift) in the process between a pair of observations would be approximated by the observations differences, so the choice of $w_{\min}$ should be driven by the expected stochasticity, and so should increase as the inter-observation time interval increases, as also observed in the three tables.

\subsection{Schl\"ogel model}
For the Schl\"ogel model, nMESA is slightly more efficient than MESA which is approximately forty times as efficient as GHS17. The particle MCMC scheme of \cite{GoliWilk15} failed to converge, but a bootstrap particle filter scheme with a large number of particles did converge, although it was over two and a half orders of magnitude less efficient than MESA and nMESA. There are several reasons for these striking results.

Firstly, for typical rate parameters, $\theta$, and large values of $X$ the rates of the two reactions involving the reservoir $A$ are extremely high. Any particle filter must simulate all the reactions that occur, a number of the same order as the sum of the four reaction rates. For Sch50, GHS17, MESA and nMESA all used the scaling and squaring algorithm (see Section \ref{sec.mxexp}), with a cost proportional to the logarithm of the of total rate; for reference, runs using used the uniformisation, which is linear in the rate, led to speed reductions by factors of $270$, $105$, and $54$ for GHS17, MESA and nMESA respectively.

The bridge of \cite{GoliWilk15} tries to drive the path for the stochastic process along an approximate straight line from the current position to the next observation. For transitions both between meta states and within the higher meta state this is an extremely poor approximation to the behaviour (see Figure \ref{fig.LVSchdata}). However, particle MCMC using a bootstrap particle filter did mix. As well as needing to simulate every single one of the reactions, the large computational cost arose from the filter needing an order of magnitude more particles than was used on the Lotka-Volterra examples.

Figure \ref{fig.LVSchdata} shows that the largest changes from one observation to the next occur during transitions from one meta state to the other, but that it is in the higher meta state that the largest expansion in region coverage from the smallest cuboid containing adjacent observations is required. To fit the latter a relatively high region number (MESA with $w_{\min}=0$) or a proposal distribution that leads to a relatively high region number (GHS17) is required, but this leads to a large statespace size as well as larger $\rho$ arising from the upper end of this large statespace. For MESA, this problem is overcome using a larger $w_{\min}$.

The biggest problem with GHS17 was the requirement for the same proposal distribution for the truncation index whatever the meta-state of the process, whereas in reality the process is likely to need a high index when in the high meta state and a low index when in the low meta state. For lower $a$ values, such as $0.95$ (not shown), small region numbers were typically proposed, the calculations were relatively cheap but the chain mixed exceedingly poorly because of the enormous (possibly infinite) variance of the logarithm of the unbiased estimator of the likelihood through the quotient term in \eqref{eqn.rand.trunc}; the chains failed to converge. To bring the variance under control, the probability of proposing larger region numbers had to be increased substantially, leading to expensive calculations because, for each inter-observation interval, the final region is larger,  and because there are typically more terms in the random truncation. Increasing $\gamma$ similarly reduces the variance of the truncation estimator of the likelihood and increases the computational effort. Furthermore, for $\gamma \ge 0.2$ with $a$ large enough for visible mixing, integer overflow (see Section \ref{sec.mxexp}) became more and more frequent since the states in the larger proposed regions led to larger rates.
Specifically, for the best run, which used $\gamma=0.2$, integer overflow occurred on $~120$ of the iterations with a maximum true $\rho\approx 1.2\times 10^{11}$ and $d$ values in excess of $3000$. For MESA with $w_{\min}=0$, integer overflow occurred on $10$ of the $10^4$ iterations, with a maximum true $\rho\approx 5.7\times 10^9$, and $d$ values up to $\approx 1400$; no overflow occurred when $w_{\min} \ge 10$. For nMESA the maximum value of $\rho d$ was $<1.1\times 10^8$.

\begin{table}
\begin{center}
  \caption{Tunings and results for GHS17, MESA and nMESA for the Sch50 dataset. $^1$The bootstrap particle filter used $1400$ particles. $^2$For GHS17, $a$ was respectively $0.99$, $0.998$ (both with $\gamma=0$) and $0.98$; the run with $a=0.998$ used only $2\times 10^4$ iterations; the  ``$-$" indicates mixing so poor that ESS could not be estimated even approximately (estimated ESS$< 20$). $^{3}$MESA with $w_{\min}=0$ was run for $10^4$ iterations. $^4$ MESA with $w_{\min}=60$ never exited region $1$.
    \label{table.Sch50}}
\begin{tabular}{l|rrr|rrr|rrrrrr}
  \multicolumn{7}{c}{}&\multicolumn{5}{c}{ESS/min}\\
  Algorithm&$\lambda$&$\gamma$&$w_{\min}$&$T$&$\alpha_{\psi}$&$\alpha_{r}$&
  $\psi_1$&$\psi_2$&$\psi_3$&$\psi_4$&$r$ or $\bar{r}$&$\log \pi$\\
  \hline
  BS$^1$&$1.2$&&&$840782$&$18.0$&&$0.30$&$0.30$&$0.25$&$0.27$&&$0.21$\\
  GHS17$^2$&$1.0$&$0.0$&&$3271$&$0.019$&&-&-&-&-&&-\\
  GHS17$^2$&$1.0$&$0.0$&&$4158$&$4.69$&&$2.58$&$2.55$&$2.43$&$2.17$&&$2.76$\\
  GHS17$^2$&$1.0$&$0.2$&&$110333$&$0.342$&&$0.01$&$0.01$&$0.01$&$0.01$&&$0.01$\\
  \hline
    MESA$^{3}$&$1.2$&$0.4$&$0$&$31182$&$27.7$&$28.9$&$1.21$&$1.28$&$1.07$&$1.04$&$0.27$&$1.01$\\
  MESA&$1.2$&$0.4$&$10$&$4879$&$27.3$&$17.7$&$85$&$86$&$80$&$74$&$158$&$85$\\
  MESA&$1.2$&$0.4$&$20$&$4596$&$27.4$&$8.8$&$91$&$91$&$81$&$83$&$158$&$78$\\
  MESA&$1.2$&$0.4$&$40$&$3648$&$27.9$&$5.8$&$116$&$114$&$114$&$110$&$196$&$104$\\
  MESA$^4$&$1.2$&$0.4$&$60$&$7912$&$29.1$&$12.4$&$58$&$58$&$53$&$53$&$*$&$43$\\
  \hline
  nMESA  &$0.8$&$0.8$&$0$&$1322$&$26.2$ &$29.9$&$120$&$140$&$55$&$81$&$47$&$102$\\
  nMESA&$0.8$&$0.2$&$20$&$1276$&$29.1$&$9.4$&$145$&$157$&$116$&$96$&$69$&$103$\\
  nMESA&$0.9$&$0.4$&$20$&$1420$&$28.8$&$5.5$&$170$&$179$&$150$&$137$&$132$&$362$\\
  nMESA&$0.9$&$0.4$&$40$&$3325$&$39.8$&$0.22$&$118$&$117$&$105$&$107$&$78$&$98$
\end{tabular}
\end{center}
\end{table}

\subsection{Autoregulatory system}
Finally, we applied nMESA to the AR50 dataset. A run of $2\times 10^5$ iterations took approximately $40$ hours
and gave a minimum (over all parameters) effective sample size of $1491$.  Tuning runs for GW15 suggested that the same number of iterations would take around $48$ days, 
so the algorithm was not run. However, alternative, appoximate inference is available via particle MCMC using the chemical Langevin equation (CLE), a stochastic differential equation (SDE) approximation to the  evolution of the spatially discrete Markov jump process \cite[e.g.][]{Wilkinson2018stochastic}. The modified diffusion bridge of \cite{DurhGall02} was used to propose paths between the observation within a particle MCMC scheme. When simulating from an approximation to the conditioned SDE using a bridge, a discretisation time step must be chosen; the larger the time step, the smaller the computational cost of each iteration. In addition to the Monte Carlo error inherent in any MCMC scheme, the CLE approach introduces error due to the approximation of the MJP by a spatially continuous process and then due to the approximation of the temporally continuous SDE by discretising time. \cite{FGS2014} observed that a coarser discretisation can lead to a premature decrease in the right tail of the posterior for some parameters, essentially because doubling a rate parameter is equivalent to doubling the inter-observation interval and keeping the parameter the same, thus effectively doubling the discretisation interval.

Using $\Delta t=0.2$  we observed a severe truncation in the right tails of the four parameters involved in reversible reactions ($\psi_1,\psi_2,\psi_5,\psi_6$) so we decreased the time step to $\Delta t=0.05$, a run which took approximately $90$ hours for $2\times 10^4$ iterations and gave a $\min$ ESS of $1404$.

\begin{figure}
\begin{center}
  \includegraphics[scale=0.7,angle=0]{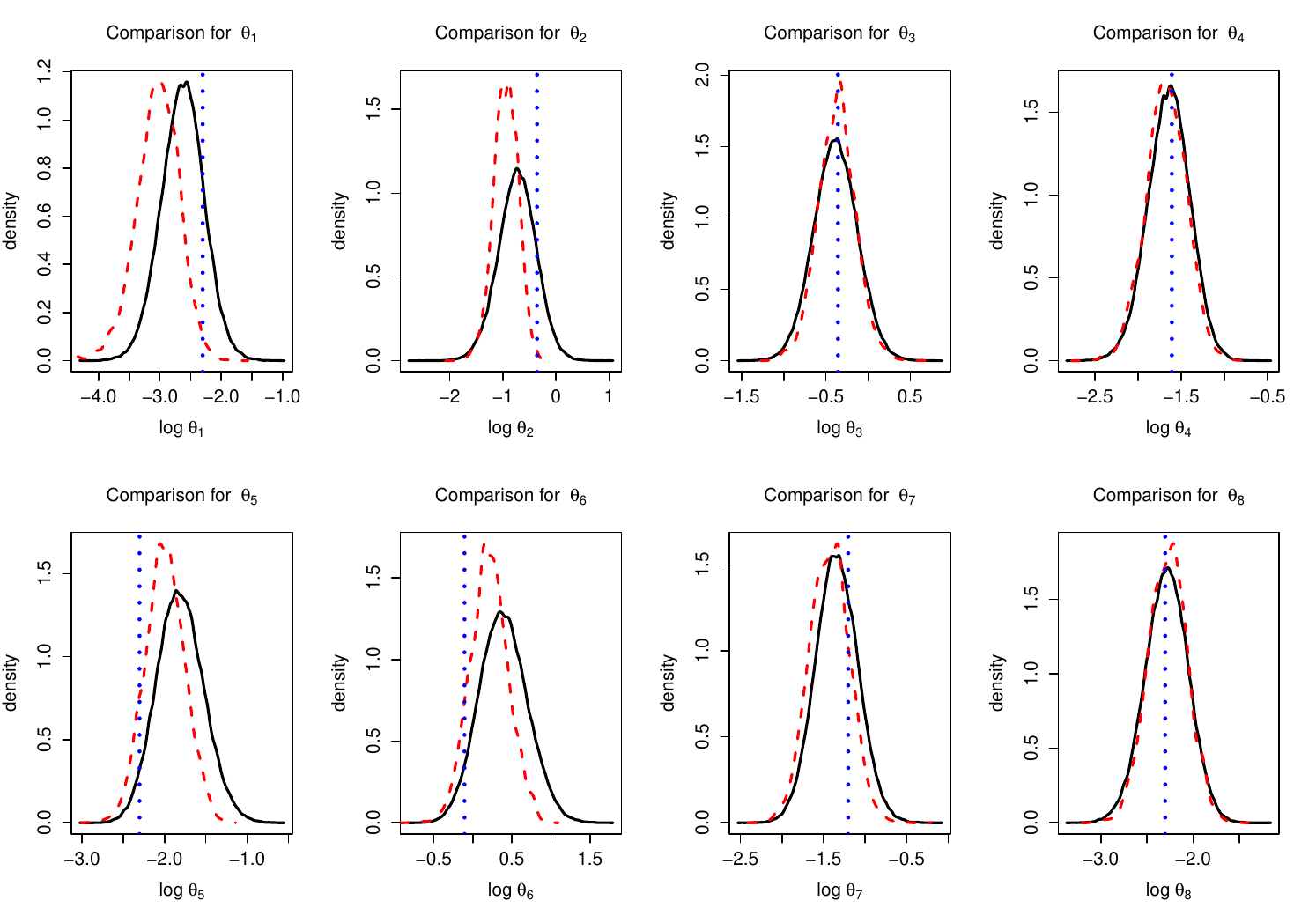}
\caption{Kernel density estimates of the parameter posteriors using nMESA (solid), and the CLE with $\Delta t=0.05$ (dashed), with the true parameter value (dotted). \label{fig.ARpostcomp}
}
\end{center}
\end{figure}

Posteriors resulting from the final discretisation are compared with the true posteriors in Figure \ref{fig.ARpostcomp}. Even with this discretisation a clear premature decay  is visible in the four parameters involved in reversible reactions. The issue is likely to be compounded for $\psi_1$, $\psi_2$ and $\psi_3$ by the error in approximating an MJP with an SDE since the first three reaction rates depend on the number of $\mathsf{DNA}$ molecules, which, with the set up detailed in Appendix \ref{sec.paramvals}, can only take values of $0$, $1$ or $2$. 

Decreasing the discretisation interval of the CLE still further would reduce (but not entirely remove) the error resulting from the CLE approximation; however, this would reduce the computational efficiency still further, and particle MCMC using the current discretisation is already only half as efficient as nMESA.

\section{Discussion}
\label{sec.discuss}
We have described the minimal extended statespace algorithm (MESA) and the nearly minimal extended statespace algorithm (nMESA) for inference on discretely and precisely observed Markov jump processes, a setting in which standard inference by particle MCMC is severely challenged. Our algorithms use the same key idea of nested regions that was used in the random-truncation algorithm of \cite{GHS17} but, in practice, are one or more orders of magnitude more efficient than that algorithm.

On the three Lotka-Volterra data sets MESA was $2.5$ - $7$ times as efficient as the best particle MCMC algorithm, and nMESA $4.5$ - $12$ times as efficient as particle MCMC. On the Schl\"ogel model, where the scaling and squaring matrix-exponentiation algorithm was used, the improvement over the best particle MCMC algorithm was over two orders of magnitude. For the autoregulatory gene model, nMESA was able to perform exact inference in a reasonable time, where no other method could. The simulations suggest that the greater flexibility of nMESA is more important for efficiency than the reduced statespace of MESA, though the advantage is only decisive for the autoregulatory model.

Both MESA and nMESA can be framed within the technique introduced in \cite{Walker2007} of rendering an infinite number of possibilities finite, and hence computable, by introducing one or more auxiliary \emph{slice} variables. Like nMESA, \cite{Walker2007} uses one auxiliary variable per observation, but a single auxiliary is also possible.
In our case, the region number bounds the dimension of the statespace of the MJP. The EA3 of \cite{Beskos2008} uses a similar set of nested regions and the idea of a minimal region containing a stochastic process, but the region is used to bound a Radon-Nikodym derivative and hence allow the exact simulation of a skeleton of a diffusion with unit volatility.  In \cite{RaoTeh2012a}, which, like MESA and nMESA, applies to continuous-time Markov chains, new states are sampled conditional on a finite set of possible event times (which are resampled in another step); if the initial condition is known and if the number of possible transitions from any state is finite then the set of possible states after finitely many jumps is also finite.

Particle MCMC and the random truncation algorithm of \cite{GHS17} are examples of pseudo-marginal MCMC \cite[]{AndrieuRoberts:2009}. Such algorithms can be viewed as introducing an extended statespace and targetting a posterior on this extended space such that the marginal, integrating out the auxiliary variables, is the target of interest. New auxiliary variables are proposed at each iteration: in the case of \cite{GHS17} these are a set of truncation variables, whereas in particle MCMC the auxiliary variables are all of the variables used by the particle filter to estimate the likelihood. Both MESA and nMESA, however, are examples of correlated pseudo-marginal algorithms \cite[]{MurrayGraham16,DelDoucPitt2018,dahlin2015accelerating} since fresh auxiliary variables are not proposed at each iteration, but instead a random walk Metropolis move from the existing variables is applied.

Other matrix exponentiation algorithms are available and, in particular, Krylov subspace-based techniques \cite[e.g.][]{SidjeStewart1999} are often used for calculating $e^A b$ for a general sparse square matrix $A$ and a vector $b$. In both the Lotka-Volterra example and the autoregulatory gene example we found this technique to be between a factor of three and an order of magnitude slower than the uniformisation technique.

Our algorithms are designed for the challenging exact-observation regime, but it would be straighforward to extend them to deal with noisy observations: nMESA via additional latent variables for the states at observation times,  MESA by considering all paths that stay entirely within $\Region_{r}$ but not $\Region_{r-1}$, and including a likelihood term at each observation time. However, as the observation noise increased, the efficiency of either algorithm would decrease gradually, whereas the efficiency of particle MCMC would increase, so that, for large-enough noise, PMCMC would be more efficient. Of more interest, is the potential for including the nested-region construction \emph{within} particle  MCMC, and this is the subject of current investigations.

\bibliographystyle{Chicago}

\bibliography{mjpexact}

\appendix

\section{Matrix Exponentiation}
\label{sec.matexpdetail}
Consider $\nu^\top \exp^{Q}=\nu^\top \sum_{i=0}^\infty Q^i/i!$, for some non-negative $d$-vector $\nu$ and $d\times d$ rate matrix $Q$. In our case for region $\Region_r$, from the $j-1$ to the $j$th observation, 
 $\nu_r$ is the $d_r+1$ vector which is $1$ at $k_r(x_{j-1})$ and $0$ everywhere else.
For a general matrix the calculation is especially difficult to evaluate efficiently yet to a pre-defined tolerance, $\epsilon$, because of the possibility of very large negative and positive numbers cancelling during the evaluation of the series; however, when $Q$ is a rate matrix this issue can be circumvented as follows.

Let $\rho=\max_{i=1,\dots,d}|Q_{i,i}|$ then $P:=I_d+Q/\rho$ has non-negative entries and is, in fact, a Markov transition matrix. Furthermore:
\[
\nu^\top e^{Q}=\nu^\top e^{\rho (P-I_d)}=e^{-\rho} \nu^\top e^{\rho P}.
\]
The \emph{uniformisation} method \cite[e.g.][]{SidjeStewart1999} evaluates  $\nu^\top e^P=\sum_{i=0}^\infty \nu^\top P^i/i!\approx \sum_{i=0}^m \nu^\top P^i/i!$, where, given $\nu^\top P^i$, and the fact that $P$ is sparse, the calculation of
$\nu^\top P^{i+1}=(\nu^\top P^i) P$ is an $\mathcal{O}(d)$ operation. The truncation point $m$ is chosen so that $\epsilon\le 1-F(m+1;\rho)$, where $F$ is the cumulative distribution function of a $\mbox{Poisson}(\rho)$ random variable, since then
\[
\nu^\top e^{Q} 1 - \widehat{\nu^\top e^{Q}} 1
=
e^{-\rho}\nu^\top \sum_{i=m+1}^\infty \frac{\rho^i}{i!}P^i 1
=\Prob{\mbox{Poisson}(\rho) \ge m+1}\le \epsilon;
\]
see \cite{sherlock2018unif} for further details.

The \emph{scaling and squaring} method \cite[e.g.][]{Dubious} uses: $e^M\equiv \left(e^{M/2^s}\right)^{2^s}$, 
for any square matrix $M$. Thus, $e^M$ can be obtained can be obtained from $e^{M/2^s}$ by squaring $s$ times. We calculate $e^{\rho P/2^s}$ using the uniformisation method \cite[but without the $\nu^\top$ term, as in][]{REIBMAN1988,Transient2018}, and revert to vector-matrix multiplications before the final squaring; see \cite{sherlock2018unif} for further details.

\section{Details of numerical experiments}
\subsection{Parameter values, process settings, and the AR50 data set}
\label{sec.paramvals}
\begin{figure}
\begin{center}
  \includegraphics[scale=0.45,angle=0]{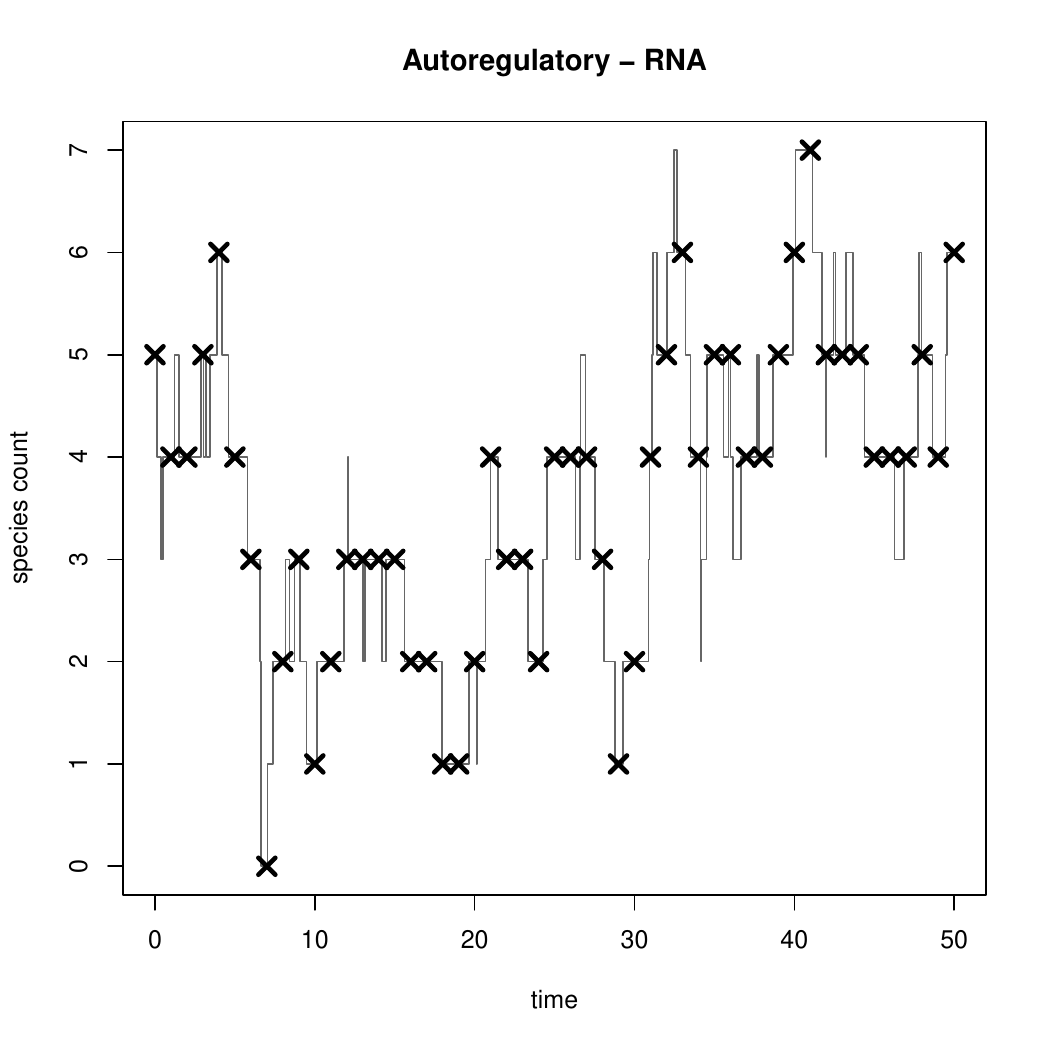}
  \includegraphics[scale=0.45,angle=0]{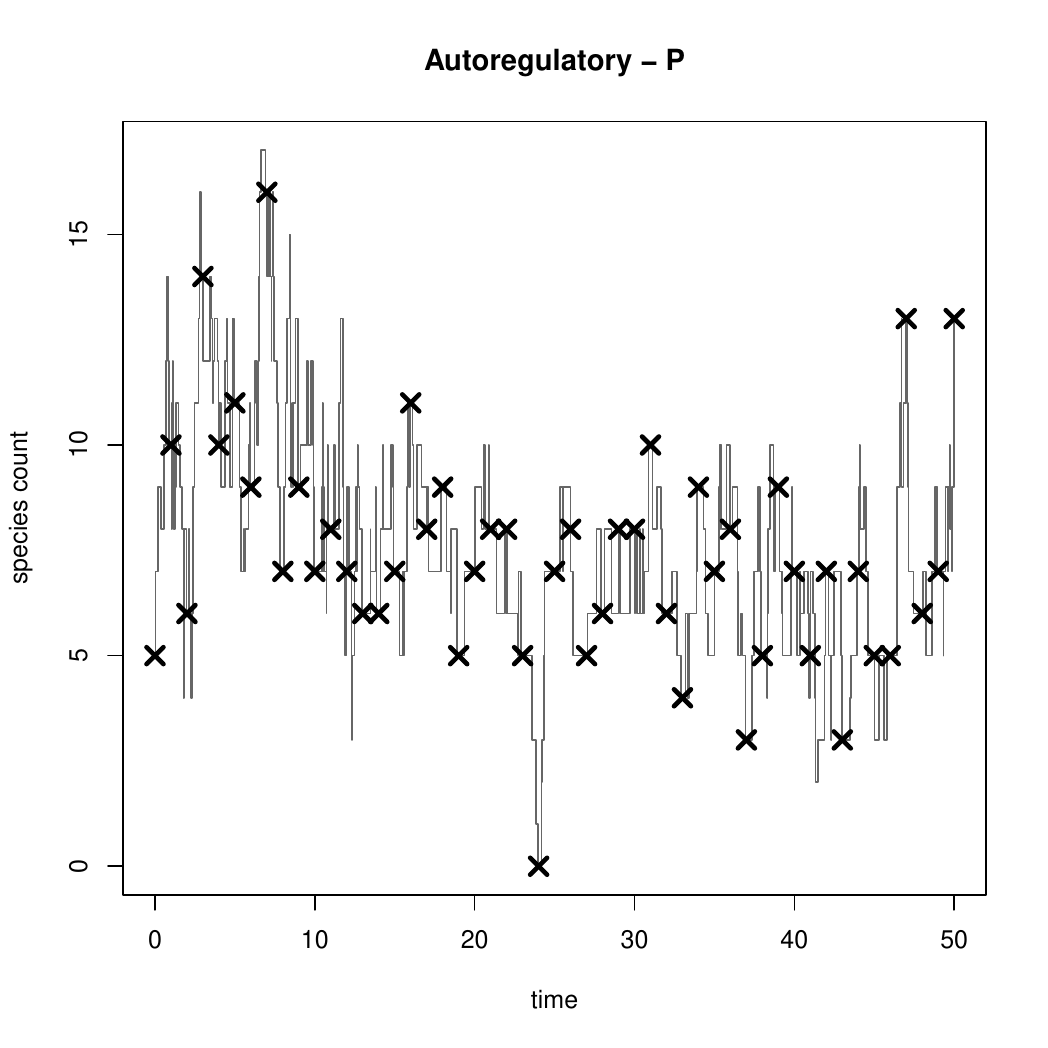}
  \includegraphics[scale=0.45,angle=0]{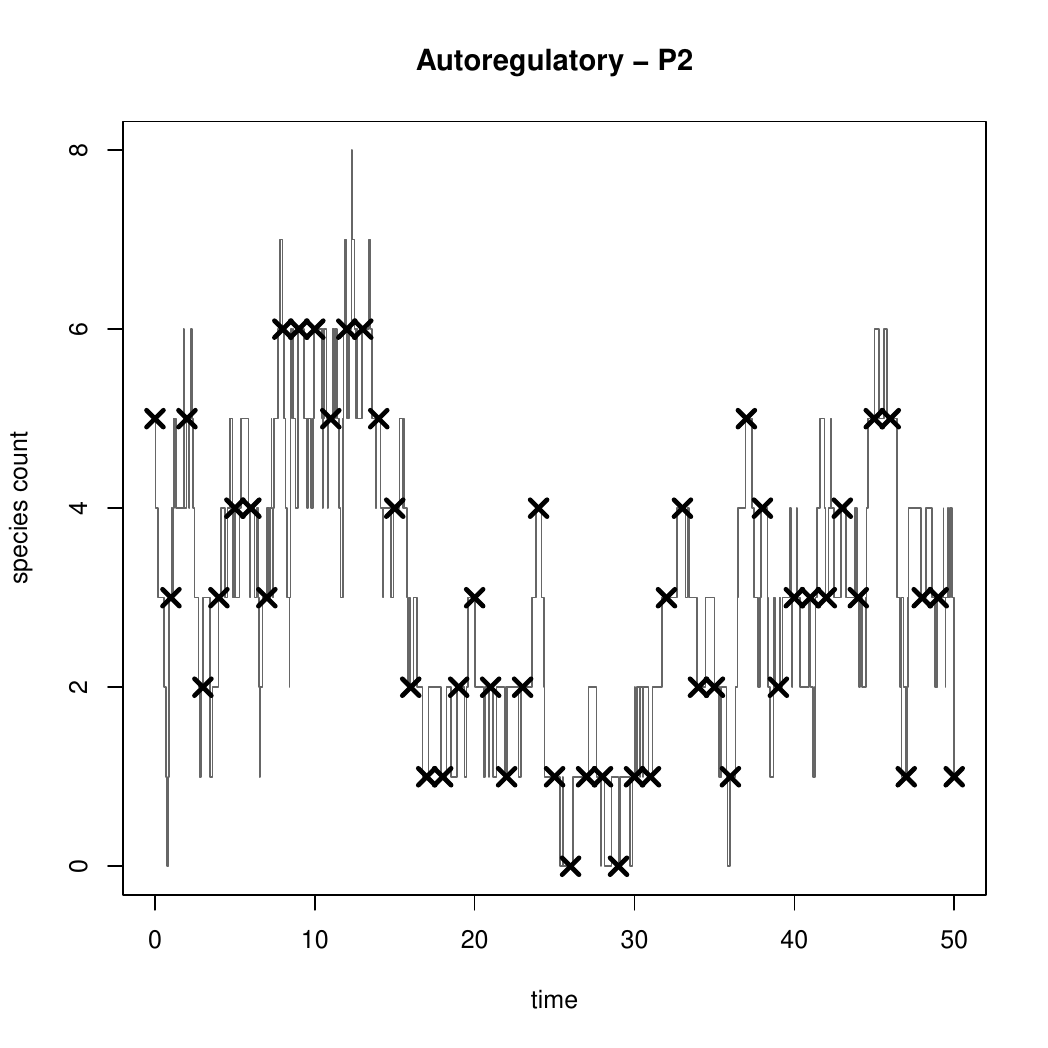}
  \includegraphics[scale=0.45,angle=0]{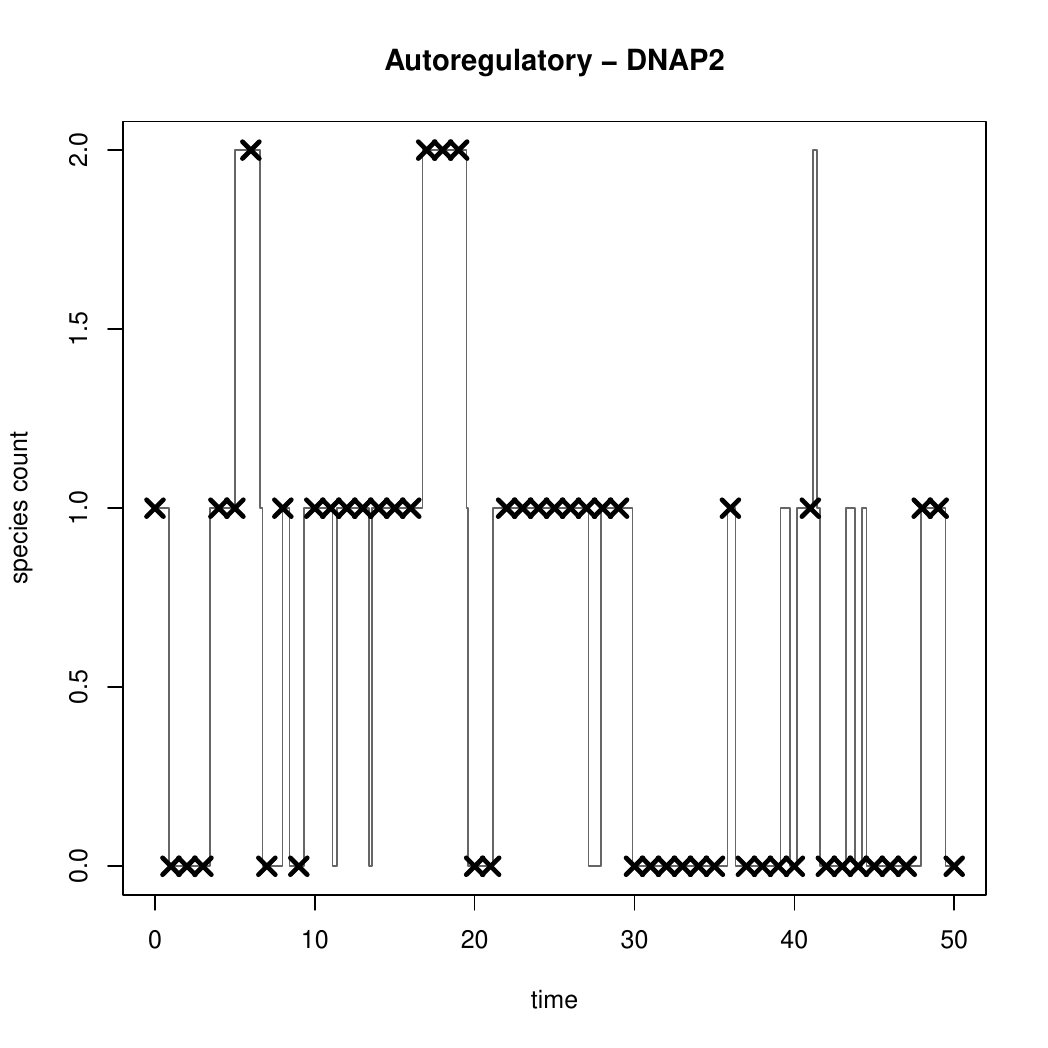}
\caption{The stochastic process that led to the AR50 dataset together with the AR50 data values: $\mathsf{RNA}$ (top left), $\mathsf{P}$ (top right), $\mathsf{P}_2$ (bottom left) and $\mathsf{DNA\cdot P_2}$ (bottom right). Quantities of $\mathsf{DNA}$ may be obtained deterministically as $DNA=2-DNA\cdot P_2$ \label{fig.ARdata}
}
\end{center}
\end{figure}

Table \ref{table.exactinfo} shows the observation and parameter information  for the five simulated data sets used in the simulation study in Section \ref{sec.numcomp}. The realisation of the autoregulatory system and the associated AR50 dataset are provided in Figure \ref{fig.ARdata}.

\begin{table}
\begin{center}
  \caption{Observation and parameter information for our five data sets of exact observations, together with the parameter values used and the abbreviated name for the data set. \label{table.exactinfo}}
\begin{tabular}{l|c|c|rrr|l}
  Process&$\theta$&$x_0$&$t_{end}$&$\Delta t$&$n_{obs}$&Name\\
  \hline
  Lotka-Volterra  &$0.3,0.4,0.01$&$(30,40)$ &$20.0$&$1.0$&$20$&LV20\\
  &&&&$0.5$&$40$&LV40\\
  &&&&$2.0$&$10$&LV10\\
  \hline
  Schl\"ogel       &$3.0,0.5,0.5,3.0$&$(0)$&$200.0$&$4.0$&$50$&Sch50\\
  \hline
  Autoregulatory   &$0.1,0.7,0.7,0.2$&
  {$(5,5,5,1,1)$}&  $25.0$&$0.5$&$50$&AR50\\
  &$0.1,0.9,0.3,0.1$&&&&
\end{tabular}
\end{center}
\end{table}

For the Lotka-Volterra model we assigned independent \emph{a priori} distributions of:
$\psi_1\sim \mathsf{N}(\log(0.2),1)$, 
$\psi_2\sim \mathsf{N}(\log(0.2),1)$ and
$\psi_3\sim \mathsf{N}(\log(0.02),1)$. The Schl\"ogel model parameters were \emph{a priori} independent with $\psi_i\sim \mathsf{N}(\log(1),1)$, $i=1,\dots,4$. For the autoregularory model, parameters were \emph{a priori} independent with
$\psi_i \sim \mathsf{N}(\log 0.2, 1)$, $i=1\dots,4,6,\dots 8$, and
$\psi_5 \sim \mathsf{N}(\log 0.2, 0.1)$. Both $\theta_5$ and $\theta_6$ describe the rates for the reversible dimerisation of $P$ and are very poorly identified by the data, although their quotient is well identified \cite[e.g.][]{GolightlyWilkinson:2005}; the tighter prior for $\psi_5$ ensures that the behaviour of the MCMC algorithm is not almost entirely dominated by this one reaction.

\subsection{Tuning GHS17}
\label{sec.tuningGHS}
The algorithm of GHS17 was tuned by first fixing $\psi$ at some sensible value (here the known true value, but in practice it would be set to the posterior mean from a training run) and recording the log-posterior, $\log \pi$, at each iteration.  For any given choice of $\gamma$, the parameter $a$ was adjusted to achieve the maximum ESS/sec for $\log \pi$. Still with $\psi$ fixed, the ESS/sec of $\log \pi$ was then investigated for different $\gamma$ at the optimal $a$ for each. In all cases it was found that $\gamma=0$ gave optimal performance. Then with the optimal $\gamma$ and $a$ parameters, $\lambda$ was adjusted to give an approximately optimal ESS.

\end{document}